\documentclass[a4paper]{aa}
\usepackage{apj2aa}
\usepackage{astron} 
\usepackage{float,epsfig,psfig}
\usepackage{pstricks}

\def\hea4{{\it HEAO~A4}}
\def\heaoa2{{\it HEAO~A2}}
\def\heao1{{\it HEAO~1}}

\def\amin{$^\prime$}

\def\eg{{\it e.g.}~}

\def\h0{$H_{\rm o}=50$~km~s$^{-1}$~Mpc$^{-1}$}
\def\q0{$q_{\rm o}$}

\def\etal    {{et~al.}~}

\def\cms3  {~{cm$^{-3}$}}

\begin{document}

  \title{Probing the Intracluster star-light with Chandra}
 \author{A.~Finoguenov\inst{1,2}, R.P.~Kudritzki\inst{3}, C.~Jones\inst{2}}

 \offprints{A. Finoguenov, alexis@xray.mpe.mpg.de}

 \institute{Max-Planck-Institut f\"ur extraterrestrische Physik,
             Giessenbachstra\ss e, 85748 Garching, Germany 
 \and
Smithsonian Astrophysical Observatory, 60 Garden st., MS 3, Cambridge,
  MA 02138, USA
  \and
 Institute for Astronomy, University of Hawaii, 2680 Woodlawn Drive,
  Honolulu, Hawaii 96822, USA}

 \date{Received October 5 2001; accepted March 20 2002}  



 \abstract{ 
 We propose a method to test for the presence of intracluster
 star-light using X-ray binaries as a trace population. We discuss a
 particular application of this method to observations of the Virgo cluster.
}

\maketitle
\section{Introduction}
The integrated light and the mass-to-light ratio in galaxy clusters plays an
important role in several aspects of observational cosmology. Carlberg \etal
(1996), for example, use the mean M/L ratio determined for the CNOC cluster
sample to derive an estimate of the matter density of the universe,
$\Omega_{\rm o}$. Mass-to-light ratios are important diagnostics of
biased/non-biased star-formation in clusters (Bahcall \etal 2000), a key
quantity for the explanation of the heavy element enrichment of the ICM
(Arnaud \etal 1992; Renzini \etal 1993), and important for describing the
statistical properties of cluster surveys (Borgani \& Guzzo 2001). The
classical approach to estimate the optical luminosity of the cluster was to
sum the light from the individual member galaxies. A detailed review on the
luminosity function of galaxies can be found in Binggeli, Sandage, Tammann
(1988). Recent studies provide more refined integrated light estimates,
adding in dwarf galaxies and an estimate for the amount of diffuse stellar
light (also called the intergalactic stellar population, IGSP), as was
recently done for the Coma cluster (Gregg \& West 1998).

The result of galaxy merging and galaxy harassment as proposed by Moore
\etal (1996) will leave stellar debris that accumulates in the cluster
potential. Therefore the observation of an intracluster stellar component
would place interesting constraints on the history of galaxy collisions in
clusters. The efficiency of tidal stripping is controlled by the probability
of a galaxy crossing the cluster core and by the duration of the impact. As
simulations show, the two effects compensate each other and the tidal radius
is similar among clusters and equals $100h^{-1}$ kpc (Klypin \etal
1999).

Observationally, the determination of the level of such diffuse emission is
very difficult due to its low optical surface brightness. It should be
observed best in nearby clusters, but there it can easily be confused with
background and foreground emission. The present sensitivity limit for
optical searches for IGSP corresponds to 20\% of the total light of the
cluster (Melnick \& Sargent 1977). This value may still be considered
uncertain, as values up to 80\% are still cited (\eg\ Feldmeier \etal
1998). One of the solutions proposed to solve this problem is the
observation of individual bright objects, that represent the light density
of the IGSP. The most important sources are supernovae, red giants and
planetary nebulae.  For the Virgo Cluster, the detection of intra-cluster
red giants (Ferguson \etal 1998) and PN (Arnaboldi \etal 1996, Mendez \etal
1997) provided evidence for the existence of an intracluster stellar
population. However recently Kudritzki \etal (2000) showed that many of the
intra-cluster PN candidates are Lyman-alpha emitting background galaxies,
although Freeman \etal (2000) demonstrated that a significant fraction must
be real PN belonging to an intracluster stellar population. Mass estimates
for this population suggest that it contains roughly 10\% to 20\% of the
stellar mass in the Virgo cluster galaxies. Durrell \etal (2002) using
independent method based on the red giant stars confirmed this conclusion.

With the advent of high-angular resolution, X-ray astronomy could play
a leading role in this seemingly purely optical field, via studies of
the intracluster X-ray binaries.

\section{The method}

Chandra observations reveal a numerous population of galactic X-ray binaries
(XRB's) in most elliptical galaxies (Sarazin, Irwin, Bregman 2000;
Finoguenov \& Jones 2001). With luminosities of $10^{37-39}$ ergs/s, these
sources are easily detected even at the distance of Virgo cluster
galaxies. The total emitted flux is proportional to the optical luminosity
of stars (Matsushita 1998) and within individual galaxies, the number
density of the XRB's follows the distribution of the stellar light
(Finoguenov \& Jones 2002; Sarazin, Irwin, Bregman 2001). Finding one XRB
with a luminosity\footnote{Spectral characteristics of XRB's are not
uniform. Thus to compare different measurements, one should specify the
spectral model assumed. We use the disk blackbody model of Makishima \etal
(1986) with the characteristic temperature of 0.5 keV and cite the
bolometric luminosity. This spectral model corresponds to the soft excess in
the XRB, which is well matched to the energy window of Chandra. If we use
the mean spectral index of 0.4 instead, the luminosity in the 0.4--10 keV is
twice as high. Spectrally harder sources are detected with Chandra at lower
countrate.}  exceeding $10^{37}$ ergs/s corresponds to revealing
$4.3\times10^8 L_{\odot}$ light in the B band. Although, the sensitivity of
the PN method is higher, $0.5\times10^8 L_{\odot}$ for each detected PN, for
bright [OIII] PN the limit $5\times10^8 L_{\odot}$ is used (Durrell \etal
2002).  Moreover, the first results for the Virgo PN survey indicate that it
is limited by contamination by background objects, rather than the lack of
detections. Therefore, we propose to use X-ray binaries, as another
indicator for the level of the intracluster stellar population.  Given the
large field of view of contemporary X-ray telescopes, finding of 1 XRB in a
single pointing corresponds to probing the average surface brightness of
intracluster star-light on the $\mu_{B}=30$ mag arcsec$^2$ level. Typical
surface brightness of the intracluster star-light is $\mu_{B}\sim27$ mag
arcsec$^2$.

At the moment, only a few Chandra studies of the X-ray binary population in
early-type galaxies have been published. Relations between the optical and
hard X-ray luminosity, based on ASCA studies alone have a scatter of a
factor of 1.5 (Matsushita 2001) for most of the galaxies, while higher X-ray
luminosities are met, they are attributed to AGN activity, in the form of
bright central point source or hard diffuse X-ray emission. White (2001)
demonstrated that the deviation from the $L_{X}-L_{B}$ relation correlates
with the specific frequency of globular clusters. This dependency could
easily be eliminated by comparison with the optical image. Our normalization
corresponds to the case of nearly (90\%) clean sample of non-globular XRB.

Theoretical modeling of the XRB's luminosity function predicts its
gradual evolution with time since star-formation (Wu 2001). Most
drastic changes are expected between the disk and the bulge population
of XRB's, as revealed by observation of M81 (Tennant \etal
2001). Therefore, having determined the luminosity function of Virgo
intracluster XRB's, we will be able to differentiate between the
scenarios for diffuse light production, which is not possible in any
other methods.

\includegraphics[width=3.2in]{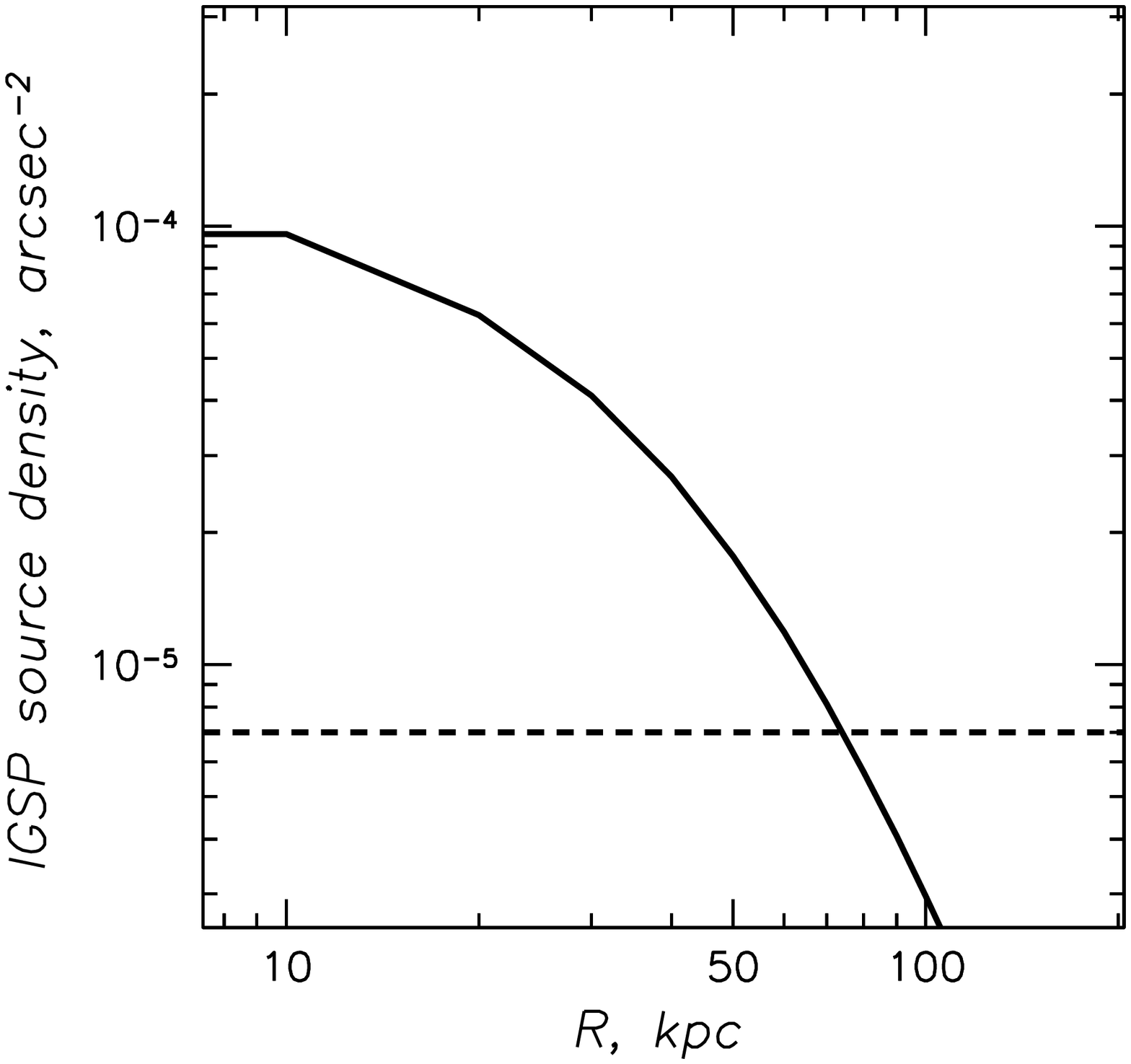}

\figcaption{Predicted X-ray source (for luminosities exceeding $10^{37}$
ergs/s) density of the Intracluster Stellar Population, having 20\% the
luminosity of Virgo cluster galaxies (solid line). The dashed line
represents 10\% of the expected contribution from CXB from Giacconi \etal
(2001). The profile is centered on the concentration of early-type galaxies
between M86 and M87, as found by Schindler, Binggeli, B\"ohringer (1999).
\label{fig:igsp}
}

The feasibility of the XRB detection in Virgo is illustrated in
Fig.\ref{fig:igsp}. We assumed that the distribution of diffuse light has a
core radius equal to the cluster tidal radius and a total amount of light in
the IGSP equal to 20\% of the total light in Virgo galaxies, according to
the revised estimate using observations of planetary nebulae (Freeman \etal
2000).

Using the model presented in Fig.\ref{fig:igsp}, detection of 1 XRB
(with luminosity exceeding $10^{37}$ ergs/s) corresponds to an IGSP
contribution of 0.2\% to the total light in the Virgo cluster. The
gradient in the source number density is expected to be sharp
(see. Fig.\ref{fig:igsp}). The amplitude of the CXB correlation
function on such angular scales (10\amin) is also $\sim10$\%
(Vikhlinin \& Forman 1995; Giacconi \etal 2001). This is therefore the
principal limitation of the method, unless optical follow-up or X-ray
spectral information is used. The resulting sensitivity is $\sim1$\%
of the total light in the Virgo cluster.  Emission from the hot X-ray
gas in the Virgo cluster, acting as a background, puts stringent
requirements on the angular resolution. For regions in the Virgo
cluster close to M87 (less than half a degree), even for Chandra,
variation of the PSF (point spread function) over the field of view
(Weisskopf \etal 1996) results in the strongly decreasing sensitivity.

Study of the spectral characteristics of the XRB in M84 have revealed that
they are spectrally different from the constituents of the CXB (Finoguenov
\& Jones 2002). Of particular interest are the binaries with a soft
component, which exhibit very similar spectra, characterized by a
multicolor disk black-body model (Makishima \etal 1986; {\it diskbb}
model in XSPEC) with a central black-body temperature of 0.5 keV. For
the luminosity range of interest ($10^{37}-10^{39}$ ergs/s), over 50\%
of the XRB in M84 exhibit such a spectrum. By limiting the sources to
only those, we can reduce substantially the contamination by CXB to
the sample and thus increase the sensitivity of this method to the
limit implied by the quality of the hardness ratio determination
(typically 30 source counts are required). For an ACIS-I exposure of
40 ksec (or ACIS-S exposure of 25 ksec), such determination will be
possible for one third of the sample, with a resulting sensitivity is
1\% of the total light in the Virgo cluster. Most of the background
objects will be type-1 QSO, for which optical counterpart can readily
be found in USNO A2.0 catalogs. Type-2 QSO do not overlap in X-ray
colors with XRBs due to characteristic strong absorption. This allows
for an independent check of the X-ray color selection.

Detection of PNs was also reported for the field far off the Virgo
center. Due to lower flux from the thermal gas the requirements on the
quality of imaging are reduced, so routine detection of XRBs in
fore-coming XMM shallow surveys, (covering large areas, typical for
X-ray surveys, challenging for optics) will be feasible.



The proposed X-ray method for estimating the amount of intracluster
star-light provides a useful alternative to optical methods and is less
affected by systematics in the search of candidates, which was recently
realized to be a problem for the PN (Planetary Nebulae) method (Kudritzki
\etal 2000). Therefore the use of X-ray observations provide an important
test of the results obtained optically. In addition, the shape of the
luminosity function of the X-ray binaries can shed light on the age of the
intracluster stellar population.

\begin{acknowledgements}

The authors thank the referee, Jimmy Irwin, for useful suggestions on the
manuscript. This work was supported by NASA grants GO0-1045X and AG5-3064
and the Smithsonian Institution.  AF has benefited from discussions with
Hans Boehringer. AF acknowledges receiving the Max-Plank-Gesellschaft
Fellowship.
\end{acknowledgements}

\bibliographystyle{aabib99}

\end{document}